\documentclass[12pt]{article}
\begin{document}
\title{CHAOS ENFORCED INSTANTON TUNNELLING IN ONE-DIMENSIONAL MODEL
WITH PERIODIC POTENTIAL}
\author{V.I. Kuvshinov, A.V. Kuzmin and R.G. Shulyakovsky \\
Institute of Physics,
National Academy of Sciences of Belarus, \\
Scarina av., 68, 220072 Minsk, BELARUS, \\
e-mails: kuvshino@dragon.bas-net.by, \\
avkuzmin@dragon.bas-net.by, \\
shul@dragon.bas-net.by}\maketitle

\begin{abstract}
The influence of chaos on properties of dilute instanton gas in
quantum mechanics is studied. We demonstrate on the example of
one-dimensional periodic potential that small perturbation leading
to chaos squeezes instanton gas and increases the rate of
instanton tunnelling.
\end{abstract}

\section{Introduction}

Semiclassical properties of classically chaotic dynamical systems
(quantum chaos \cite{QuantumChaos}) is a reach rapidly developing
field of research. One of the attractive phenomenon obtained in
this direction is chaos assisted tunnelling (\cite{ChaosTunn} and
references therein).

In this work we analytically investigate influence of the
perturbation leading to chaos on the rate of instanton~\cite{BPST}
transitions on the example of one-dimensional periodic potential.
Instantons represent alternative way to describe quantum
tunnelling and play an important role not only in quantum
mechanics (recent example \cite{Jap}), but mainly in the modern
theories of particle physics, e.g. electroweak theory, QCD, SUSY,
etc. (for review see~\cite{Shuryak}). The search of instantons,
initiated in DESY, is one of targets of modern QCD~\cite{RSh}. The
main problem in this direction is an extremely small probability
of instanton-induced events. It is of importance to study if chaos
can increase probability of instanton tunnelling. We demonstrate
on the particular example of simple quantum mechanical system that
it is really possible.

Hamiltonian of the system under consideration is taken in the form

\begin{equation}\label{1}
\tilde{H} = \frac{1}{2}\tilde{p}^{2} + \omega_{0}^{2}\cos{x} -
\epsilon x \sum_{n=-\infty}^{+\infty} \delta(t - n\tilde{T}),
\end{equation}
$\tilde{T}$ is the real time period of perturbation. The systems
with spatially periodic potential are well-studied in solid-state
physics \cite{kittel} and instanton physics \cite{Rajar}.
Perturbation used in (\ref{1}) was widely exploited in the systems
exhibiting quantum chaos \cite{Physica}.

For applying instanton technique we have to consider classical
solutions of Hamilton equations in {\it imaginary} (Euclidian)
time. Hamiltonian (\ref{1}) has the same form (translated on
$\pi$) in Euclidian time as in real one.

In Euclidian time classical Hamiltonian of the system looks as
follows $H=H_0+V$,
\begin{equation}\label{H0}
H_{0} = \frac{1}{2}p^{2} - \omega_{0}^{2}\cos{x},
\end{equation}
\begin{equation}\label{perturbation}
  V = \epsilon x \sum_{n=-\infty}^{+\infty}
  \delta(\tau-nT).
\end{equation}
Here $H_{0}$ is non-perturbed Hamiltonian of the system and $V$ is
the Euclidian potential of the perturbation. In these expressions
variables without tilde denote corresponding quantities in
imaginary time.

\section{Instanton gas in non-perturbed system}
Let us consider at first non-perturbed Hamiltonian (\ref{1}), i.e.
put $\epsilon =0$. The system is characterized by degenerated
vacuum structure on the classical level:
\begin{equation}
x^{vac}_n(t)\equiv |n>=\pi n, \quad n=\pm 1,\pm 3,...
\end{equation}
As is well-known any tunnelling transition in this system can be
represented in terms of path integral in imaginary time:
\begin{equation}\label{r1}
<m|e^{-H_0\Gamma}|n>=\int [Dx]e^{-S[x]},
\end{equation}
where $\Gamma$ is a time of the transition between wells, $S[x]$
denotes Euclidean action.
The main contribution to (\ref{r1}) is given by instantons.
One-instanton configurations are classical solutions of Euclidean
equations of motion and describe tunnelling between neighboring
vacua. They can be easy found as well as one-instanton action
(see, for example~\cite{Rajar})
\begin{equation}\label{I}
x^{inst}(\tau,\tau_0)=\pm\left(
arctg\left[e^{\omega_0(\tau-\tau_0)}\right]-\pi\right),\,
S[x^{inst}(\tau,\tau_0)]\equiv S^{inst}=8\omega_0,
\end{equation}
where arbitrary parameter $\tau_0$ is a center of the instanton;
sings $'+'$ and $'-'$ correspond to instantons and
anti-instantons.

Multi-instantons are not exact classical solutions, but they give
leading contribution to the amplitude of tunnelling (\ref{r1})
between distant wells
\begin{equation}\label{multi}
x^{(n)}(\tau)=\sum_{i=1}^{n}x^{inst}(\tau,\tau_{i}), \qquad
S[x^{(n)}(\tau)]=nS^{inst}.
\end{equation}
Here we suppose that time intervals between centers of single
instantons $\tau_i$ are not too close to each other (dilute
instanton gas approximation)~\cite{Rajar}.

Let us focus our attention on the amplitude $A_q(\Gamma)$ of $q$
one-instanton transitions during time interval $\Gamma$. The
difference between instantons and anti-instantons is not essential
in these calculations. Using standard instanton
technique~\cite{Rajar} it is easy to obtain Poisson distribution
on $q$ for the amplitude in Gauss approximation:
\begin{equation}
A_q(\Gamma)=N\frac{1}{q!}e^{-qS^{inst}}\left(\Gamma\sqrt{S^{inst}}\right)^q,
\end{equation}
where factor $N$ provides correct normalization, $A_0(\Gamma)$
means absence of instantons. The average number of instantons for
the time $\Gamma$ is
\begin{equation}\label{<q>}
<q>=e^{-S^{inst}}\Gamma\sqrt{S^{inst}}.
\end{equation}
Thus in imaginary time we have a gas consisted of instantons
(\ref{I}) with average time interval between them $\eta_0$ and
average density $\rho_0$:
\begin{equation}\label{density}
\eta_0=\frac{\Gamma}{<q>}=\frac{e^{S^{inst}}}{\sqrt{S^{inst}}},\qquad
\rho_0=\eta_0^{-1}=e^{-S^{inst}}\sqrt{S^{inst}}=
2\sqrt{2\omega_0}e^{-8\omega_0}.
\end{equation}
Here, as usual, we neglected by the instanton size.

Let us suppose large value of one-instanton action (\ref{I}). It
corresponds to high energy barriers. In this case we obtain
strongly rarefied instanton gas.

It should be noted that result (\ref{density}) was obtained by
using of both exact (\ref{I}) and approximate (\ref{multi})
classical solutions. Taking into account only exact solutions
leads to zero instanton density $\rho_0$ (or infinite interval
$\eta_0$). Finite density can appear in non-perturbed system
(\ref{1}) only due to approximate solutions contribution.

\section{Squeezing of instanton gas due to small perturbation}
Now we consider the case with $\epsilon \neq 0$ and estimate an
average interval between instanton transitions (inverse  density
of instantons in instanton gas) for the perturbed system. For this
purpose we represent the perturbation (\ref{perturbation}) in the
form
\begin{equation}\label{newpert}
V= \frac{\epsilon}{T} x \left(2\sum_{m=-\infty}^{+\infty}
\cos{(m\nu \tau)} + 1\right).
\end{equation}
Here $\nu \equiv 2\pi / T$.

Perturbation (\ref{newpert}) destroys separatrix of non-perturbed
system (\ref{H0}) and on its place stochastic layer appears
\cite{Lihtenberg}. We primarily estimate the width of stochastic
layer. We use method reviewed in \cite{Sagdeev}. We apply it to
the system (\ref{H0}-\ref{perturbation}) which was not considered
in that works, although non-perturbed Hamiltonian (\ref{H0}) and
perturbation (\ref{perturbation}) were used independently from
each other.

Exact equation of motion for action variable is
\begin{equation}\label{eq1}
  \dot{I} = \frac{d I}{d H_{0}} \dot{H_{0}} = - \frac{\epsilon}{\omega T} \dot{x}\left(
 2\sum_{m=0}^{+\infty} \cos{(m\nu \tau)} + 1\right),
\end{equation}
Here $I$ denotes action variable and $\omega \equiv d H_{0}/d I$
is a frequency of nonlinear oscillations (see \cite{Lihtenberg}).

Consider behavior of the system near separatrix (in imaginary
time). Dependence of velocity $\dot{x}$ on time has the form of
rare soliton-like impulses. Each impulse corresponds to rapid
transition between two neighbor peaks of potential. Long time
interval between two impulses is the time to get over a peak. It
is seen from equation (\ref{eq1}) that action variable changes
mainly during the impulse of velocity. Introduce phase of external
force $\varphi$ defined as $ \dot{\varphi} = \nu$.

To study chaotic behavior of the system we transform differential
equation (\ref{eq1}) to discrete mapping
\begin{equation}\label{map}
\left\{ \bigskip
\begin{array}{c}
  \overline{I} = I + \frac{\epsilon}{T \omega(I)} C(I, \varphi) \medskip \\
  \overline{\varphi} = \varphi + \frac{\pi \nu}{\omega (\overline{I})},
\end{array}
\right.
\end{equation}
where ($\overline{I},\overline{\varphi}$) denote values of action
variable and phase just after the impulse of velocity,
($I,\varphi$) are the same quantities after previous impulse and
\begin{equation}
  C(I, \varphi) = - \int_{\Delta \tau} d \tau \dot{x} \left( 2\sum_{m=0}^{+\infty}
  \cos{(m\varphi(t))} + 1 \right).
\end{equation}
Here we integrate over time interval of velocity's impulse $\Delta
\tau$.  From (\ref{map}) we obtain (for small $\epsilon$)
\begin{equation}
  \overline{\varphi} \simeq \varphi + \frac{\pi \nu}{\omega (I)} - \frac{ \epsilon \nu^{2}}{
   2\omega^{3}}\frac{d\omega}{dI} C(I, \varphi).
\end{equation}
Therefore parameter of local instability (defined in
\cite{Sagdeev}) is
\begin{equation}\label{K}
  K = \left| \frac{\delta \overline{\varphi}}{\delta \varphi} -1 \right| = \frac{\pi
  \epsilon\nu}{T \omega^{3}}\left|\frac{d\omega}{dI}\right| C_{0}, \quad  C_{0}\equiv
  \left|\frac{\partial C}{\partial \varphi}\right|
\end{equation}
In the vicinity of separatrix $C_{0}$ can be calculated
\begin{equation}\label{CO}
  C_{0}\simeq 4\pi \frac{\sinh{(\pi \nu/2\omega_{0})}}{\cosh^{2}{(\pi \nu/2\omega_{0})}},
\end{equation}
Using (\ref{CO}) and estimation for $|d\omega / dI|$ at $|H-
H_{s}|\ll H_{s}$
 we can represent the parameter of local instability (\ref{K}) in the
form
\begin{equation}\label{Kfin}
  K\simeq \frac{2 \epsilon \nu^{2}}{\omega_{0}}\frac{\sinh{(\pi
  \nu/2\omega_{0})}}{\cosh^{2}{(\pi \nu/2\omega_{0})}} \frac{1}{|H-H_{s}|}.
\end{equation}
Here $H_{s}\equiv \omega_{0}^{2}$ is the energy of non-perturbed
system on the separatrix.

Condition $K\geq 1$ means that dynamics of the system is locally
unstable. Local instability leads to mixing and chaos
\cite{Sagdeev}. Thus condition $K=1$ gives the estimation for the
width of the stochastic layer as follows
\begin{equation}\label{width}
  |H_{b} - H_{s}|\approx \frac{4\epsilon \nu^{2}}{\omega_{0}}e^{-\pi \nu /2\omega_{0}},
\end{equation}
under the assumption that $\nu > \omega_{0}$. Here $H_{b}$ is an
estimated value of energy on boundary of the stochastic layer. A
set of trajectories appearing in the stochastic layer can be
considered as new multi-instanton-like exact solutions of
Euclidian equations of motion for perturbed system. These
solutions demonstrate finite intervals between instanton
transitions. The reason is that for majority of trajectories in
stochastic layer finite time is needed to pass from one potential
well to another. Thus density of instanton gas strongly increases
in comparison with density at $\epsilon =0$.

Average time interval between instantons in the instanton gas for
perturbed system can be estimated in the following way
\begin{equation}
\eta\approx \frac{\pi}{\omega_{av}}.
\end{equation}
Here $\omega_{av}\equiv \omega([H_{b}+H_{s}]/2)$. Near the
separatrix following approximation for the frequency can be used
\cite{Sagdeev}
\begin{equation}
\omega (H)\equiv \frac{d H_{0}}{d I} \approx
\frac{\pi}{\sqrt{2}}\frac{\sqrt{H_{s} + H}}{\ln{\frac{16(H_{s} +
H)}{H-H_{s}}}}, \quad |H-H_{s}| \ll H_{s}.
\end{equation}
Thus interval between instanton transitions is the next
\begin{equation}\label{etap}
  \eta\approx  \frac{\pi \nu}{2 \omega_{0}^{2}} +
  \frac{1}{\omega_{0}}\ln{\frac{8 \omega_{0}^{3}}{\epsilon \nu^{2}}},
\end{equation}
and the density of instantons in instanton gas for the perturbed
system can be estimated as follows
\begin{equation}\label{rho}
  \rho = \frac{1}{\eta} = \left(\frac{\pi \nu}{2 \omega_{0}^{2}} +
  \frac{1}{\omega_{0}}\ln{\frac{8 \omega_{0}^{3}}{\epsilon \nu^{2}}} \right)^{-1}.
\end{equation}
We see that, in contrary to the case of non-perturbed system, for
perturbed system the density of instanton gas is large if we take
into account only exact solutions of Euclidian equations of
motion. Consideration of approximate solutions only increases the
density of instanton gas and does not lead to qualitative changes.
Comparing of densities for perturbed $\rho$ (\ref{rho}) and
non-perturbed $\rho_0$ (\ref{density}) systems gives
\begin{equation}\label{result}
  \frac{\rho}{\rho_{0}} \approx \frac{1}{\sqrt{2}} \frac{\omega_{0} \sqrt{\omega_0}
  e^{8\omega_{0}}}{\pi \nu + 2
  \omega_{0}\ln{\frac{8 \omega_{0}^{3}}{\epsilon \nu^{2}}}}.
\end{equation}
This ratio is large at large enough $\omega_0$ (or large enough
one-instanton action) and small but nonzero $\epsilon$. Thus we
obtain that small perturbation (\ref{perturbation}) can strongly
increase the density of instanton gas. For calculation of $\rho$
only exact classical solutions of Euclidian equations of motion
were taken into account. While to get non-zero value of $\rho_{0}$
we had to consider approximate solutions. Thus the limit $\epsilon
\rightarrow 0$ can not be applied directly in (\ref{result}).
Nevertheless we have correspondence between results obtained for
perturbed and non-perturbed systems in this limit. Namely, $\rho$
(see (\ref{rho})) at $\epsilon = 0$ and $\rho_{0}$ are equal to
zero if we take into account only exact solutions of Euclidian
equations of motion for {\it both} cases.


Let us apply now our formal computation to the physical phenomena
in real time. Classical instanton solutions in imaginary time
(\ref{I}-\ref{multi}) describe quantum tunneling transitions in
real time. Some observables can be directly expressed through the
instanton density $\rho$ \cite{Rajar}. In particular, rate of the
tunnelling between neighbour potential wells (the number of
tunnelling transitions per unit of time) and probability of
tunnelling are proportional to squared density of instantons
$\rho^2$. Spectrum and width of the lowest energy zone $\Delta E$
read
\begin{equation}\label{spectrum}
E_{\theta}\approx\frac{1}{2}\omega_0- 2\rho\cos{\theta},\quad
0\leq\theta\leq\pi, \qquad  \Delta E\approx 4\rho.
\end{equation}
Thus squeezing of the instanton gas and increase of the density
$\rho$ in imaginary time mean that small perturbation
(\ref{perturbation}) leading to chaos can essentially enhance the
tunnelling rate and lead to the widening of the energy zone in
comparison with non-perturbed system (\ref{H0}). Both these
results are consequences of the perturbation leading to
destruction of the single non-perturbed instanton solution and
appearance of manyfold of chaotic perturbed instantons. An
increase of the number of instanton solutions provides a larger
number of variants for particle to reach one vacuum from another
that results in the increase of the rate of tunnelling. The
decrease of the life-time of the particle in the certain vacuum of
the system means the widening of the energy zone that is obtained
in (\ref{spectrum}).

On the other hand our results can be considered as a demonstration
on the simple model of application of the instanton method to the
problem of chaos assisted tunnelling.


\section{Conclusion}
In this work we have demonstrated on the example of
one-dimensional periodic potential that small perturbation leading
to chaotic behavior of the system strongly influences on the
properties of instanton gas. Our estimations show that classical
chaos can greatly increase the density of instanton gas and rate
of instanton tunnelling.

Both instanton solutions and chaotic behavior can exist in complex
systems like field theories. The relation between chaos and
instantons in such theories is not trivial. Theory of strong
interactions (QCD) is the most interesting example, where the
investigation of instanton gas (or instanton liquid) could shed
light on the structure of hadrons~\cite{Shuryak},~\cite{CDG}.

\end{document}